\documentclass[a4paper,twocolumn]{article}
\usepackage{amsmath}
\usepackage{epsfig,subfigure}
\usepackage{cite}
\date{}
\begin{document}
\title{
\Large{
\textbf{Phase-matched second-harmonic generation in a
ferroelectric liquid crystal waveguide
}
}
}\normalsize
\author{
Valentina S. U. Fazio$^{1,3}$, Sven T. Lagerwall$^{1}$, 
Philippe Busson$^{2}$, Anders
Hult$^{2}$, Hubert Motschmann$^{3}$\footnote{e-mail:
motschma@mpikg-golm.mpg.de}   \\
$^{1}$\textit{\small{\it
Department of Microelectronics \& Nanoscience, Liquid Crystal Physics,}}\\
\textit{\small{\it
Chalmers University of Technology \&
G\"oteborg University, SE-41296 G\"oteborg, Sweden}} \\
$^{2}$\textit{\small{\it
Department of Polymer Technology, Royal Institute of Technology,
SE -10044 Stockholm, Sweden}}\\
$^{3}$\textit{\small{\it
Max-Plank-Institute of Colloids and Interfaces,
D-14424 Golm/Potsdam, Germany}}
}
\normalsize
\normalsize
\maketitle
\noindent
\textbf{Abstract}\\
\noindent
True phase-matched second-harmonic generation in a waveguide of
crosslinkable ferroelectric liquid crystals is demonstrated.
These materials allow the formation of macroscopically polar
structures whose order can be frozen by photopolymerization. 
Homeotropic alignment was chosen which offers decisive
advantages compared to other geometries. 
All parameters contributing to the conversion
efficiency are maximized by deliberately controlling the supramolecular 
arrangement.
The system has the potential to achieve practical level of performances
as a frequency doubler for low power laser diodes. 
\vspace{10mm}

\noindent
PACS number(s):\\ 
61.30.Gd (orienatational order of liquid crystals; 
electric and magnetic field effects on order) \\
42.65.Tg (optical solitons; nonlinear waveguides)\\
42.79.Nv (optical frequency converters)\\

\vspace{10mm}

\noindent
The classical domain of nonlinear optical (NLO) devices based on second-order
effects ($\chi^{(2)}$-effects) is frequency doubling which is important
for extending the frequency range of laser light sources\cite{Koechner}.
The major goals of devices based on third-order nonlinear optical
effects ($\chi^{(3)}$-effects) is the realization of optical switches as
the decisive hurdle on the way to an all-optical data processing
\cite{PrasadWilliams}.
The design concepts exploit the intensity dependent refractive index due to
$\chi^{(3)}$-interactions in Mach-Zehnder type interferometers
\cite{WalRosClaSha91a,WalRosClaSha91b}.
Recently it has been demonstrated that an
intensity dependent refractive index can also be obtained by a cascading
of second order nonlinear processes\cite{DesHagSheSte92,BaeSchSteAss98}.
This route is far more efficient than the one using $\chi^{(3)}$-effects
with currently available materials.
As a result, switching occurs at lower intensity levels\cite{Bos96}.

The figure of merit of both, frequency doublers and optical switches, is
given by the ratio of the susceptibility $\chi^{(2)}$ and refractive index
$n$ as ${\chi^{(2)}}^{2}/n^{3}$.
Organic materials possess refractive indices $n \approx 1.5$ and
thus have an edge to most inorganic materials with refractive index
$n \approx 2.2$.
Furthermore, organic molecules can be tailored according to the demands and
different desired functionalities  can be incorporated within a single
molecule \cite{Ulman}.
The inherent potential has been early recognized and meanwhile there is a
sound knowledge of the correlation between molecular structure and
corresponding hyperpolarizability, $\beta$ \cite{HanKimCho99,KucJanKaa99}.
Organic chromophores possess a remarkably high hyperpolarizability and the 
major obstacle towards efficient devices is not the availability of
suitable chromophores, but the fabrication of proper macroscopic structures.

A  high conversion efficiency requires the simultaneous maximization of
many parameters and quite often there is a trade-off between some properties. 
A crucial quantity is $\chi^{(2)}$ which is, subject to certain simplifying
assumptions, proportional to the number density of the NLO chromophores 
and to the orientational average of the hyperpolarizabilities 
\cite{MotPenArmEnz93}. 
Hence, a chromophore with high hyperpolarizability should be arranged in a
noncentrosymmetric fashion with high number density and high degree of
orientational order. 
To achieve this, mainly two concepts have been pursued so far: 
Langmuir-Blodgett (LB) films\cite{Petty, Ulman} and poled polymers 
\cite{PrasadWilliams}. 
However, due to intrinsic peculiarities of both techniques, the
chromophore is rather diluted and furthermore the films possess limited
thermal and mechanical stability. 
In this study we pursued a different strategy based on ferroelectric
liquid crystals (FLCs).

Liquid crystals (LCs) in general form highly ordered phases which 
possess an intrinsic quadrupolar order but not a dipolar one \cite{deGennes}. 
Hence for $\chi^{(2)}$ applications conventional LCs are not of any use. 
However, the picture changed with the advent of FLCs whose molecular 
symmetry allows a local dipole perpendicular to the director
\cite{MeyLieStrKel75}. 
The arrangement can be manipulated by electric field and huge single
domains can be formed. 
At this stage the orientational order within the monomeric system is
still fragile and also sensitive to slight changes in temperature. 
To overcome these problems the FLCs are further functionalized with
photoreactive groups. 
Subsequent photopolymerization leads to the formation of stable polymer 
networks \cite{TroOrrSahGed96, TroSahGedHul96, HerRudLagKom98, 
LinHerOrtArn98, FazLagZauSch99} where the polar order is frozen
(pyroelectric polymer, PP).
Various aspects of the preparation process as well as some nonlinear
optical properties are described in a recently submitted publication
\cite{FazLagZauSch99}. 
In this contribution we focus on the problem of phase-matching in
waveguide geometry and demonstrate that true phase-matching is possible 
to achieve. 

\begin{figure}
\begin{center}
\epsfig{file=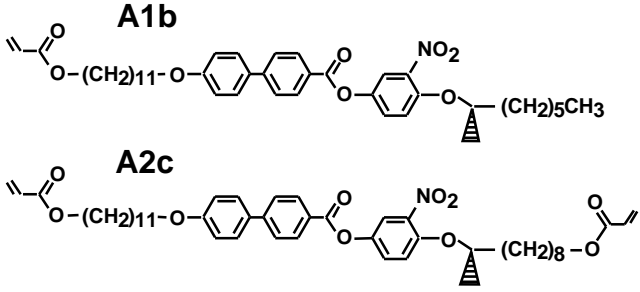, width=0.47\textwidth}
\caption{
\label{molecules}
\small{
FLC monomers used in this experiment. }
}
\end{center}
\end{figure}

\begin{figure}
\begin{center}
\epsfig{file=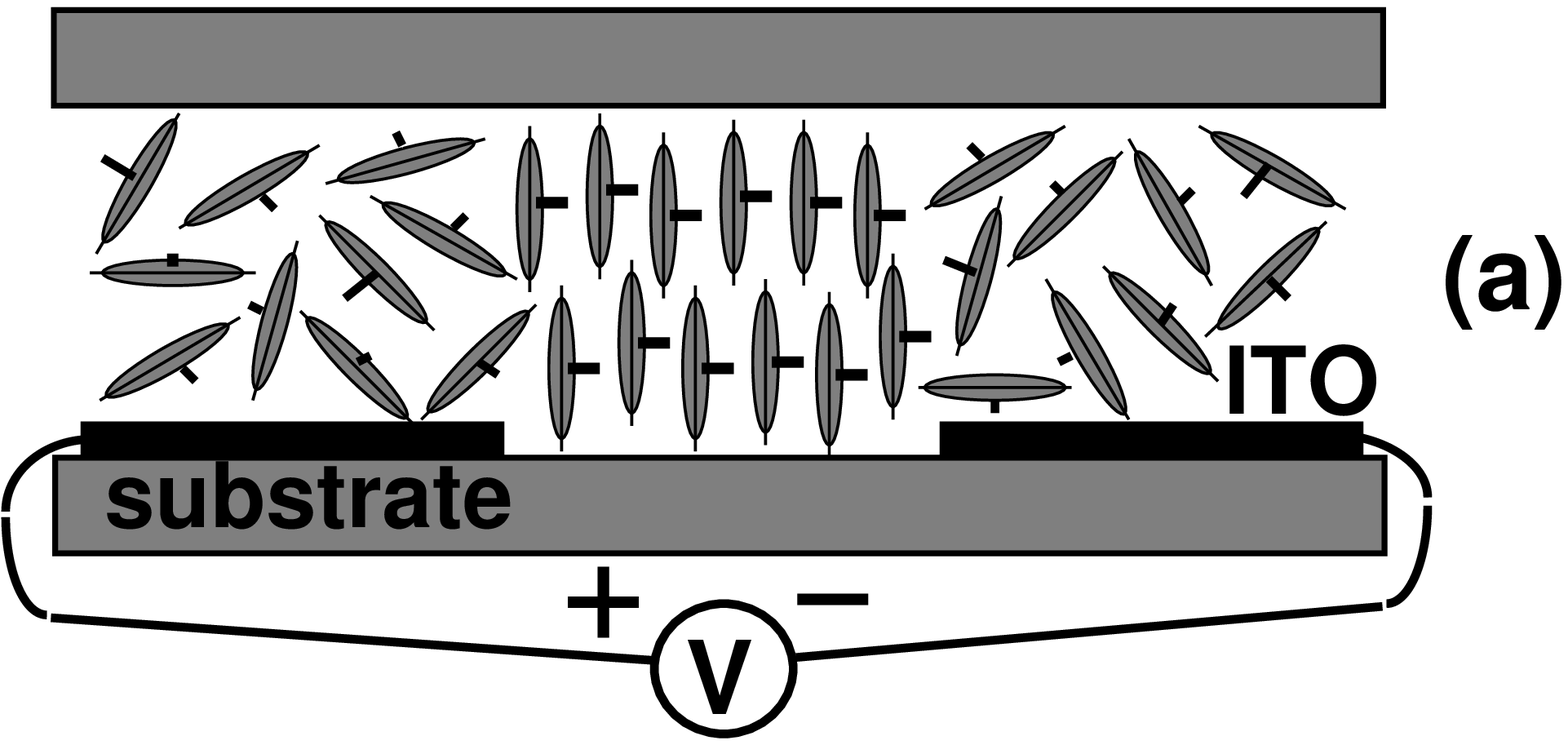, width=0.30\textwidth}\\
\vspace{3mm}
\epsfig{file=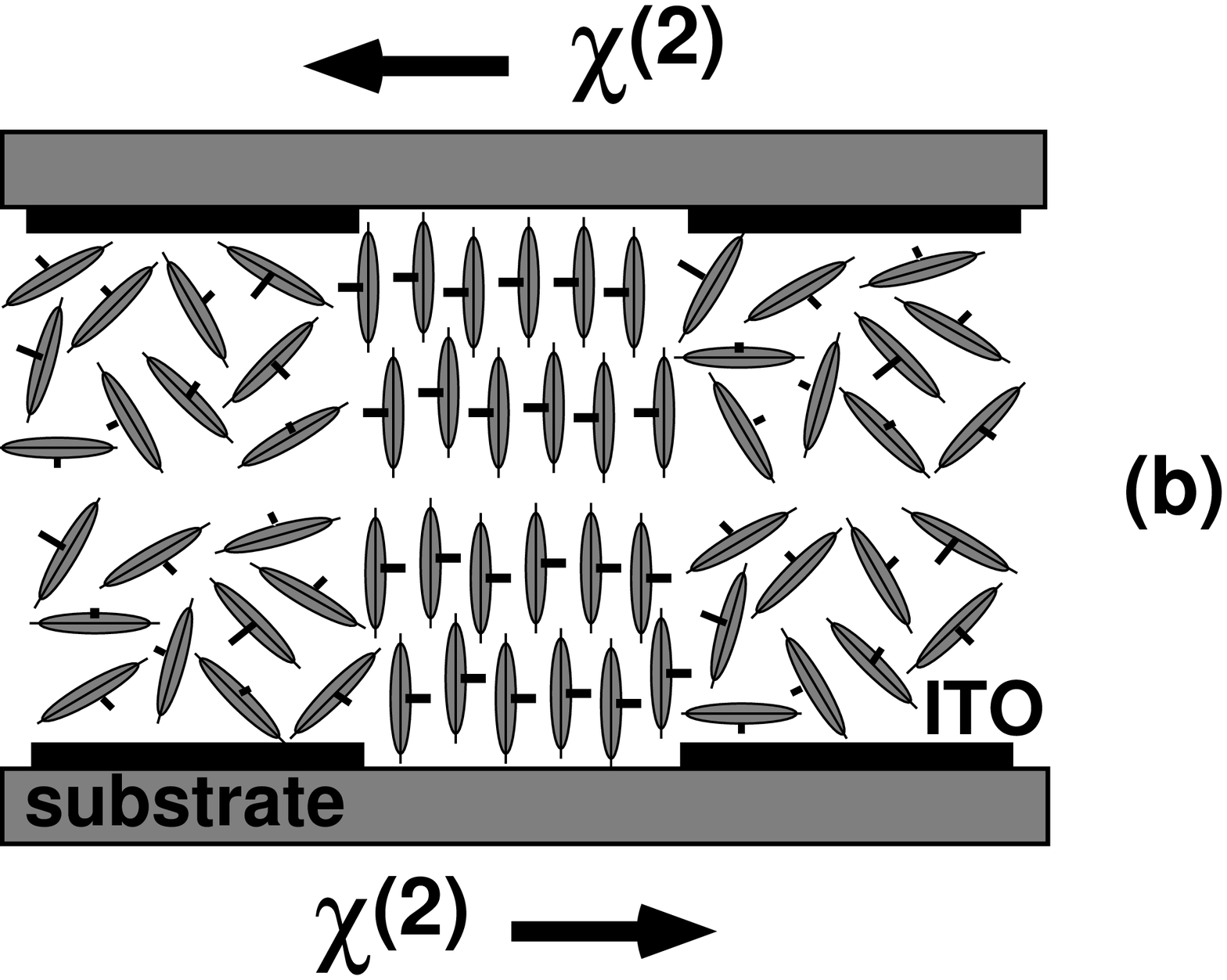, width=0.30\textwidth}
\caption{
\label{cross_sec_TOT}
\small{
(a) Cross section of a cell. Cell thickness is 540\,nm. Electrode 
distance is 100\,$\mu$m.
(b) Scheme of the $\chi^{(2)}$-inverted structure.
$\chi^{(2)}$ undergoes an abrupt sign reversal at half the total thickness
of the cell. Cell thickness is 2 $\times$ 540\,nm.}
}
\end{center}
\end{figure}

The chemical structures of the FLC monomers are shown in
Fig.~\ref{molecules}.
A mixture of 60 {\%} \textbf{A1b} and 40 {\%} \textbf{A2c}
is used which adopts at room temperature a chiral smectic C
phase\cite{LinHerOrtArn98}.
This mixture is filled in a cell depicted in Figure \ref{cross_sec_TOT}(a)
\cite{FazLagZauSch99}.
The bottom plate is equipped with parallel ITO electrodes stripes 
to achieve a quasi-homeotropic alignment: the smectic layers are aligned 
parallel to the glass plates, the molecular dipole moments
are oriented by the electric field, and the helical structure of the chiral 
phase is unwound. 
The mixture of \textbf{A1b} and \textbf{A2c} balances the trade-off
between a high polarization on one hand and the field strength
required for a manipulation of the helix on the other hand\cite{FazLagZauSch99}.
Even a moderate electric field strength is sufficient to obtain a highly
ordered structure and no aligning layers are required. 
The achieved polar order of the monomeric FLC system is then permanently fixed by
photopolymerization leading to a mechanically and thermally stable PP network.  
Without any additional preparation step this arrangement is also a channel 
waveguide for TM modes\cite{FazLagZauSch99}. 
A linear and nonlinear optical characterization is presented in
ref.\cite{FazLagZauSch99}. 
According to the prevailing symmetry, second-harmonic generation (SHG)
can only occur for $\text{TE}^{\omega}$--$\text{TE}^{2\omega}$ and
$\text{TM}^{\omega}$--$\text{TE}^{2\omega}$ modes\cite{HermannDavid97}. 
The measured nonlinear optical constants are remarkably large (up to 
1.26\,pm\,V$^{-1}$).

Phase-matching can be achieved by taking advantage of the modal
dispersion of the waveguide \cite{SteSto89}. 
The effective refractive index $n_{\text{eff}}$ of a mode is a function of 
waveguide thickness and polarization.
Thus, phase-matching requires the fabrication of a waveguide of a 
precisely defined thickness given by the linear optical constants. 
The tolerances are quite tight and already minor deviations within the 
nanometer range change the characteristics of a device.
Also, due to the dispersion of the refractive index, phase-matching is
only possible between modes of different order.
However, even if this is achieved, the resulting efficiency may still 
be rather low due to the small value of the overlap integral of
the electric field distribution of the interacting modes across
the cross-sectional area\cite{SteSto89}
\begin{equation*}
\mathcal{I} = \int_{0}^{\infty} 
\frac{{\chi^{(2)}}_{ijk}}{{\chi^{(2)}}_{\text{eff}}}
{{E_{i}}^{(m^{\prime},\omega)}}(z) \, {{E_{j}}^{(m^{\prime},\omega)}}(z) \,
{{E_{k}}^{(m, 2\omega)}}(z) \, dz,
\end{equation*}
where ${\chi^{(2)}}_{ijk}$ is the second-order susceptibility tensor,
${\chi^{(2)}}_{\text{eff}}$ is the effective second-order susceptibility,
and ${E_{i}}^{(m^{\prime},\omega)}(z)$ is the electric field distribution 
of the $m^{\prime}$-th mode of frequency $\omega$ across the waveguide 
thickness.
Field distributions of modes of different order yield a nearly vanishing 
overlap integral and a poor conversion efficiency \cite{SteSto89}.
A way out of this dilemma is to influence the susceptibility tensor
\cite{PenMotArmEze94}. 
A reversal of sign of $\chi^{(2)}$ at the nodal plane 
of  the electric field
distribution of the first-order mode maximizes the value of the overlap
integral and thus enables a phase-matching scheme
${\text{TM}_{0}}^{\omega}$--${\text{TE}_{1}}^{2\omega}$ and
${\text{TE}_{0}}^{\omega}$--${\text{TE}_{1}}^{2\omega}$.
The sign of $\chi^{(2)}$ can be reversed by reversing the polar
order of the chromophores.

\begin{figure}[t]
\begin{center}
\epsfig{file=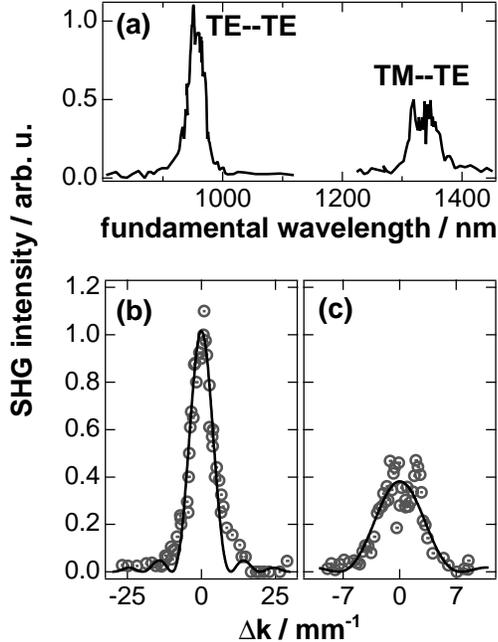, width=0.41\textwidth}
\caption{
\label{PM_norm1}
\small{
(a) Phase-matching peaks as a function of the fundamental light
wavelength.
(b) ${\text{TE}_{0}}^{\omega}$--${\text{TE}_{0}}^{2\omega}$
phase-matching peak at 955\,nm, data and fit.
(c) ${\text{TM}_{0}}^{\omega}$--${\text{TE}_{0}}^{2\omega}$
phase-matching peak at 1337\,nm, data and fit.}
}
\end{center}
\end{figure}

The desired inverted waveguide structure can be fabricated using
the sandwich geometry shown in Figure \ref{cross_sec_TOT}(a).
The top plate of a 540\,nm thick cell was removed.
No damage occurred in this preparation process (the mean roughness
is on the order of few nanometers\cite{FazLagZauSch99} as confirmed by 
atomic force microscopy). 
The bottom plate with the polymer network was cut in two pieces of
equal size ($\approx$ 4\,mm) and the parts were glued onto each
other with inverse polarities in the channel region  as
illustrated in Figure~\ref{cross_sec_TOT}(b). 
Waveguide modes were excited by \textit{end-fire} coupling.
The second-harmonic (SH) light was collected at the end of 
the guide and measured as a function of the fundamental light 
wavelength with a photomultiplier.
A quadratic dependence of the SH light intensity on the fundamental 
one was established to ensure the true nature of the observed signal.
The linear constants and the thickness of the waveguide were measured prior
to the experiment and used to predict the wavelengths at which 
phase-matching occurs.
According to these data and with a total cell thickness of
2\,$\times$\,540\,nm, TE--TE phase-matching should occur at 958\,nm
and TM--TE phase-matching at 1311\,nm.
Indeed, the experiment confirms these predictions: TE--TE phase-matching
was observed at 955\,nm and TM--TE at 1337\,nm, as shown in Figure
\ref{PM_norm1}(a).

The width of the peaks in Figure \ref{PM_norm1}(b) and (c) depends on
the known dispersion of the refractive indices and on the interaction
length $L$ in which fundamental and second harmonic light are in phase.
The interaction length can be determined by a fit of the experimental
data to the function
\begin{equation*}
I_{2\omega} \propto \text{sinc}^{2} \left(\frac{L \, \Delta k}{2} \right),
\label{sinc}
\end{equation*}
where $\Delta k =4 \pi
[n_{\text{eff}}(2\omega)-n_{\text{eff}}(\omega)]/\lambda_{\omega}$,
with $L$ as the only unknown parameter.
Figures \ref{PM_norm1}(b) and (c) present the experimental data together
with the corresponding fits.
The interaction lengths are listed in Table~\ref{table}.
\begin{table}
\begin{center}
\caption{
\label{table}
Interaction lengths and SH conversion efficiencies for the two
phase-matching peaks in Figure~\protect\ref{PM_norm1}.
}
\begin{tabular}{|c|c|c|}
\hline
phase-matched modes & $L$ [mm] & $\eta$ [{\%}
W$^{-1}$ cm$^{-2}$]\\
\hline
\hline 
${\text{TE}_{0}}^{\omega}-{\text{TE}_{1}}^{2\omega}$ & 
0.59\,$\pm$\,0.08 & 0.05\,$\pm$\,0.02 \\
\hline
${\text{TM}_{0}}^{\omega}-{\text{TE}_{1}}^{2\omega}$ &
0.82\,$\pm$\,0.15 & 0.26\,$\pm$\,0.06 \\
\hline
\end{tabular}
\end{center}
\end{table}
In a sample without $\chi^{(2)}$-inversion the SH signal was about 1000
times smaller than that of an inverted sample at the phase-matching
condition, which demonstrated the superior performance due to the
optimization of the overlap integral in our geometry.
The conversion efficienciency 
\begin{equation*}
\eta=\frac{P_{2\omega}}{{P_{\omega}}^{2} \, L^{2} }
\end{equation*}
of the two phase-matching schemes is also given in Table~\ref{table}.
The values are among the largest ones in organic materials.
Also, the confinement of TM modes three-dimensionally in the waveguide
yields a larger conversion efficiency for TM--TE than for TE--TE 
phase-matching.  

The demonstration of true phase matching in a waveguide format using 
FLCs
is a major step towards a more general use of these materials  for
NLO devices.  
FLCs maximize the possible number density of active chromophores and
this, together with a high degree of orientation, leads to remarkably
high values of the off-resonant nonlinear susceptibilities. 
Phase-matching was achieved between modes of different order using the modal
dispersion of the waveguide. 
The concept of an inverted structure maximizes the overlap integral and
thus enables high efficiency in the desired phase matching-scheme. 
We have successfully manufactured a macroscopic inverted 
waveguide and demonstrated phase-matching. 
The quasi-homeotropic alignment avoids the use of aligning layers and leads to
an inherent channel waveguide for TM modes without any additional preparation
steps, yielding a very high conversion efficiency for TM--TE
phase-matching scheme.
Another major feature is that the order of the monomeric FLC
is made permanent by photopolymerization. 
The photopolymerization does not lead to any degradation of the quality
of the waveguide, as it is for instance observed in LB 
films\cite{Ulman}.
Apparently the intrinsic fluidity of FLC heals all distortions caused by
the formation of new bonds. 
The polar network is thermally and mechanically stable and all samples
kept their NLO properties over the monitored period of several months.
Thus, the system has the potential to achieve practical levels of
performance. \\

\noindent
The authors are grateful to Dr. S. Schrader for helpful
discussions and to Prof. H. M\"ohwald for generous support and encouraging
discussions.
V. S. U. Fazio and S. T. Lagerwall are grateful to the TMR European Programme
(contract number ERBFMNICT983023) and to the Swedish Fundation for Strategic
Research for financial support.
P. Busson acknowledges the financial support from the Swedish Research 
Council for Engineering Science (TFR, grant 95-807).

\bibliography{journal2,let14}

\begin{thebibliography}{10}

\bibitem{Koechner}
W.~{Koechner}.
\newblock {\em Solid state laser engineering}.
\newblock {Berlin} {Springer}, 1992.

\bibitem{PrasadWilliams}
P.~N. {Prasad} and D.~J. {Williams}.
\newblock {\em Introduction to nonlinear optical effects in molecules and
  polymers}.
\newblock {John} {Wiley} \& {Sons}, 1991.

\bibitem{WalRosClaSha91a}
D.~M. {Walba}, M.~B. {Ross}, A.~A. {Clark}, R.~{Shao}, K.~M. {Johnson}, M.~G.
  {Robinson}, J.~Y. {Liu}, and J.~{Dorowski}.
\newblock {\em Mol. Crys. Liq. Crys.}, 198:51, 1991.

\bibitem{WalRosClaSha91b}
D.~M. {Walba}, M.~B. {Ross}, A.~A. {Clark}, R.~{Shao}, K.~M. {Johnson}, M.~G.
  {Robinson}, J.~Y. {Liu}, and J.~{Dorowski}.
\newblock {\em J. Am. Chem. Soc.}, 113:5472, 1991.

\bibitem{DesHagSheSte92}
R.~{De} {Salvo}, D.~J. {Hagan}, M.~{Sheik}-{Bahae}, G.~I. {Stegeman},
  E.~W.~{Van} {Stryland}, and H.~{Van} {Herzeele}.
\newblock {\em Opt. Lett.}, 17(1):28, 1992.

\bibitem{BaeSchSteAss98}
Y.~{Baeck}, R.~{Schiek}, G.~I. {Stegeman}, G.~{Assanto}, and W.~{Sohler}.
\newblock {\em Appl. Phys. Lett.}, 72(26):3405, 1998.

\bibitem{Bos96}
C.~{Bosshard}.
\newblock {\em Adv. Mater.}, 8:385, 1996.

\bibitem{Ulman}
A.~{Ulman}.
\newblock {\em An introduction to ultrathin organic films: from
  {Langmuir}-{Blodgett} to self-assembly}.
\newblock {Academic} {Press} {Boston}, 1991.

\bibitem{HanKimCho99}
S.~{Hahn}, D.~{Kim}, and M.~{Cho}.
\newblock {\em J. Phys. Chem}, 103(39):8221, 1999.

\bibitem{KucJanKaa99}
S.~{Kucharski}, R.~{Janik}, and P.~{Kaats}.
\newblock {\em J. Mater. Chem.}, 9(2):395, 1999.

\bibitem{MotPenArmEnz93}
H.~{Motschmann}, T.~{Penner}, N.~{Armstrong}, and M.~{Enzenyilimba}.
\newblock {\em J. Phys. Chem}, 97(??):3933, 1993.

\bibitem{Petty}
M.~C. {Petty}.
\newblock {\em {Langmuir}-{Blodgett} films: an introduction}.
\newblock {Cambridge} {University} {Press}, 1996.

\bibitem{deGennes}
P.~G. de~{Gennes}.
\newblock {\em The physics of liquid crystals}.
\newblock {Oxford} {University} {Press}, 1974.

\bibitem{MeyLieStrKel75}
R.~B. {Meyer}, L.~{Liebert}, L.~{Strzeleki}, and P.~{Keller}.
\newblock {\em J. Physique}, 36:L69, 1975.

\bibitem{TroOrrSahGed96}
M.~{Trolls{\aa}s}, C.~{Orrenius}, F.~{Salh{\'e}n}, U.~W. {Gedde}, T.~{Norin},
  A.~{Hult}, D.~{Hermann}, P.~{Rudquist}, L.~{Komitov}, S.~T. {Lagerwall}, and
  J.~{Lindstr\"om}.
\newblock {\em J. Am. Chem. Soc.}, 118:8542, 1996.

\bibitem{TroSahGedHul96}
M.~{Trolls{\aa}s}, F.~{Sahl\'en}, U.~W. {Gedde}, A.~{Hult}, D.~{Hermann},
  P.~{Rudquist}, L.~{Komitov}, S.~T. {Lagerwall}, and B.~{Stebler}.
\newblock {\em Macromol.}, 29(7):2590, 1996.

\bibitem{HerRudLagKom98}
D.~S. {Hermann}, P.~{Rudquist}, S.~T. {Lagerwall}, L.~{Komitov}, B.~{Stebler},
  M.~{Lindgren}, M.~{Trolls{\aa}s}, F.~{Sahl\'en}, A.~{Hult}, U.~W. {Gedde},
  C.~{Orrenius}, and T.~{Norin}.
\newblock {\em Liq. Crys.}, 24(2):295, 1998.

\bibitem{LinHerOrtArn98}
M.~{Lindgren}, D.~S. {Hermann}, J.~{\"Ortegen}, P.-O. {Arntzen}, U.~W. {Gedde},
  A.~{Hult}, L.~{Komitov}, S.~T. {Lagerwall}, P.~{Rudquist}, B.~{Stebler},
  F.~{Sahl\'en}, and M.~{Trolls{\aa}s}.
\newblock {\em J. Opt. Soc. Am. B}, 15(2):914, 1998.

\bibitem{FazLagZauSch99}
V.~S.~U. {Fazio}, S.~T. {Lagerwall}, V.~{Zauls}, S.~{Schrader}, P.~{Busson},
  A.~{Hult}, and H.~{Motschmann}.
\newblock {\em submitted to \protect\textit{Phys. Rew. E}}, 1999.

\bibitem{HermannDavid97}
David~Sparre {Hermann}.
\newblock {\em Interaction of light with liquid crystals}.
\newblock PhD thesis, {G\"oteborg} {University} and {Chalmers} {University} of
  {Technology}, 1997.

\bibitem{SteSto89}
G.~{Stegeman} and R.~{Stolen}.
\newblock {\em J. Opt. Soc. Am. B}, 6(4):652, 1989.

\bibitem{PenMotArmEze94}
T.~L. {Penner}, H.~R. {Motschmann}, N.~J. {Armstrong}, M.~C. {Ezenyilimba}, and
  D.~J. {Williams}.
\newblock {\em Nature}, 367:49, 1994.

\end{thebibliography}
\end{document}